\documentclass[12pt]{article}
\usepackage{amssymb}
\usepackage{amsfonts}
\usepackage{epsfig}
\usepackage{graphicx}
\usepackage{epsfig}
\usepackage{xcolor}

\begin{document}

\title{The specific shapes of gender imbalance in scientific authorships: a
network approach}
\author{{\small Tanya Ara\'{u}jo$^{a,b}$ and Elsa Fontainha$^{a}$} \and {\small $^{a}$ISEG
(Lisbon School of Economics \& Management) Universidade de Lisboa, } \and
{\small Rua do Quelhas, 6 1200-781 Lisboa Portugal} \and {\small $^{b}$Research
Unit on Complexity and Economics (UECE)} \and {\small Rua Miguel Lupi, 20
1249-078 Lisboa Portugal}}
\date{}

\maketitle

\begin{abstract}
{\small Gender differences in collaborative research have received little
attention when compared with the growing importance that women hold in
academia and research. Unsurprisingly, most of bibliometric databases have a
strong lack of directly available information by gender. Although
empirical-based network approaches are often used in the study of research
collaboration, the studies about the influence of gender dissimilarities on
the resulting topological outcomes are still scarce. Here, networks of
scientific subjects are used to characterize patterns that might be
associated to five categories of authorships which were built based on
gender. We find enough evidence that gender imbalance in scientific
authorships brings a peculiar trait to the networks induced from papers
published in Web of Science (WoS) indexed journals of Economics
over the period 2010-2015 and having at least one author affiliated to a
Portuguese institution. Our results show the emergence of a specific pattern
when the network of co-occurring subjects is induced from a set of papers
exclusively authored by men. Such a male-exclusive authorship condition is
found to be the solely responsible for the emergence that particular shape
in the network structure. This peculiar trait might facilitate future
network analyses of research collaboration and interdisciplinarity.}
\end{abstract}

Keywords: co-occurrence networks, gender, research collaboration,
interdisciplinarity, bibliometrics, minimum spanning tree

\section{Introduction}
The handiness of powerful computational instruments and recent improvements
in multidisciplinary methods are providing researchers an ever-greater
opportunity to investigate societies in their complex nature \cite{Ban12}.
Several research outcomes have been showing that men and women differ in
characteristics that could be related to their collaboration patterns.
Research collaboration is increasing in frequency and scope. It is driven,
among other causes, by growing relationship across scientific disciplines,
improvement of the efficiency in research resources in projects and
development of information and communication technologies \cite{Abramo15}.
The motivations \cite{Beaver01}, strategies, patterns and impacts on
scientific productivity in quantity and quality in research collaboration
have received great scholarly attention (\cite{Boz04};\cite{Cainelli15};\cite%
{Ductor15}). The patterns vary across space (\cite{Hoek10};\cite{Stef01}),
academic ranks \cite{Abramo14}, professional origins \cite{Beaver78} and
scientific disciplines \cite{Tsai16}.

Economic science makes connections with many other scientific disciplines,
like Statistics or Social Sciences, like Sociology, History or Management (%
\cite{Kric06};\cite{Pie02}). Economics shows a growing increase of
co-authorship (\cite{Cainelli15};\cite{Bar88};\cite{McDo83}). On average, a
researcher in Economics had less than one co-author in the 1970s, 1.24
co-authors in the 1980s and 1.67 in 1990s (\cite{Goy06};\cite{Tsai16}).

Gender differences in collaborative research concerning motivations,
strategies, patterns and impacts on science performance have received little
attention, contrasting with the growing importance that women hold in
academia and research. The literature shows mixed results about the gender
differences concerning research collaboration strategies \cite{McDo83},
impacts (\cite{McDo83};\cite{Abramo15};\cite{Meng16}-\cite{Ror15}) and
patterns (\cite{Abramo13}-\cite{Uhly15}).

Bibliometric studies and survey analysis are the main methodologies to the
study of research collaboration \cite{Bara02}. Large bibliometric databases
like Web of Science (\cite{Adri13};\cite{Har16};\cite{Sug13}) are the main
sources used to bibliometric analysis. However, that bibliometric databases
have a strong weakness concerning the study of the differences by gender;
they do not include information separated by male-female and the way to
overcame that weakness is to obtain the information from the first name \cite%
{Nal04} or the family name of the author \cite{Kol15}.

The present paper seeks to build upon the previous analysis about gender
aspects in research collaboration which literature was recently surveyed in
\cite{Abramo13}. Here, we intend to contribute to at least two points of the
literature: the differences of research collaboration and interdisciplinary
participation by gender. Focusing in Economics, a scientific subject
strongly connected to other scientific domains \cite{Pie02} and constructing
five categories of articles in a gender authorship perspective, this study
addresses both issues: research collaboration and interdisciplinarity.

Applying a network approach and using as unit of analysis articles indexed
in the Web of Science (WoS) this analysis maps the research
collaboration by gender within dozen of scientific subjects, all associated
with Economics. The choice of network approaches to study research
collaboration in economics \cite{Kric06} has been extensively embraced. It
often relies on the discovery of patterns of collaborations within
researcher communities, aiming to find the influence of individual
researchers in the networks using citation analysis. Reference \cite%
{Beaver78}, in the first complete theory of scientific collaboration, list
and discuss the causes for that collaboration. They stress that it is
necessary, when scientists deal with research questions, that cross
disciplinary bounds. They also identify a large variation in collaboration
by discipline, which is being further investigated in more recently
published studies (\cite{Abramo13};\cite{Boz11}).

Its well known that the adoption of a network approach allows the modeling
of social structures from a bottom-up perspective, as resulting from the
interaction (or likeness) of individual characteristics \cite{Ban12}.
Moreover, as the individual characterization might be driven by multiple
aggregate concerns, the network approaches allow for simultaneously
considering that multiplicity of individual aspects and the consequences of
the aggregate structures themselves on the emergence of collective patterns.
Meanwhile, in the adoption of a network approach, one shall be aware that
the choice of a given network representation is only one out of several
other ways to look at a given set of elements. As connecting the elementary
units of a system may be conceived in many different ways, that choice may
depend strongly on the available empirical data and on the questions that a
network analysis aims to address \cite{Arau16}.

The main question addressed in this paper is whether some relevant
characteristics of research collaboration would emerge in networks where
subjects are linked whenever they co-occur in a common paper. We
hypothesized that gender imbalance in authorship of papers might influence
the shape of those networks, allowing to uncover patterns from gender
differences. If it happens, the emerging patterns may help to understand
important characteristics of research collaboration, of the relationship
among subjects and its relation to gender.

The paper is organized as follows: next section presents the empirical data
we work with and some preliminary statistical results. Section 3 describes
the network approach and the results from its application. Section 4
concludes.

\section{\protect\bigskip The Data}

The Web of Science (WoS) is one of the major bibliometric databases
(together with Google Scholar and Scopus) and includes all scientific
subjects. It comprises a total of 11,990 Journals (8,778 from Science and
3,212 from Social Sciences) \cite{Har16}.  Concerning the scientific domain
of Economics, it includes 334 Publications\footnote{%
In Journal of Citation Report 2016 the number of journals is 344.}. The WoS
classifies each journal in one or more subjects (or categories).

Taking as examples the journals \textbf{Journal of Informetrics} and \textbf{Research Policy}, the former is classified in
"Computer Science, Interdisciplinary Applications" and "Information Science \& Library Science", while the latter is classified in "Management" and "Planning \& Development".

From the original WoS database a selection of articles was carried on
adopting as criteria: articles published in WoS indexed journals over the
period 2010-2015, having Economics as scientific subject and at least one
author affiliated to a Portuguese institution.

Our motivation to focus on the field of Economics and on the papers whose
authors are affiliated to Portuguese institutions is twofold:

\begin{enumerate}
\item Economic science makes connections with many other scientific subjects.

\item According OECD data, Portugal presents the highest percentage of women
in research during the period of 2004-2012 (OECD, 2016).
\end{enumerate}

Consequently, our approach is applied to a data set comprising 1,138 papers
published in 2010, 2011, 2012, 2013, 2014 and 2015 and having Economics as
the main subject matter.

Besides Economics, each paper may have extra (or secondary) subjects. Table
1 presents the set of secondary (extra) subjects found in our data set. Each
paper in the data set is coded by a string that informs about the presence
of extra subjects. In the broader set of 1,138 papers having Economics as
the main subject matter, 29 different extra (or secondary) subjects were
found.

\begin{center}
\begin{tabular}{|l|l|}
\hline
{\small Subject} & {\small Subject} \\ \hline
{\small 1 Agricultural Economics} & {\small 2 Area Studies} \\ \hline
{\small 3 Business} & {\small 4 Cultural Studies} \\ \hline
{\small 5 Environmental Science} & {\small 6 Education} \\ \hline
{\small 7 Ecology} & {\small 8 Finance} \\ \hline
{\small 9 Geography} & {\small 10 Health Policy} \\ \hline
{\small 11 History Of S.Sciences} & {\small 12 Hospitality} \\ \hline
{\small 13 Industrial Rel. \& Labor} & {\small 14 Interdisciplinary St.} \\
\hline
{\small 15 International Relations} & {\small 16 Leisure, Sport \& Tourism}
\\ \hline
{\small 17 Management } & {\small 18 Mathematics} \\ \hline
{\small 19 Occupational Health} & {\small 20 Operations Research} \\ \hline
{\small 21 Planning \& Development} & {\small 22 Political Science} \\ \hline
{\small 23 Science \& Technology} & {\small 24 Social Sciences} \\ \hline
{\small 25 Sociology} & {\small 26\ Statistics \& Probability} \\ \hline
{\small 27 Transportation} & {\small 28 Urban Studies} \\ \hline
{\small 29 Engeeniring} &  \\ \hline
\end{tabular}

{\small Table 1: Secondary (or extra) subjects besides Economics.}
\end{center}

The structure presented in Table 2 exemplifies the way we represent the
presence (and thus the co-occurrence) of subjects in each paper, it also
shows the way we organize information on gender authorship \footnote{%
The gender of the authors was identified by the first given name, because in
Portuguese, the first given name defines the gender without any ambiguity.
When the authors did not have Portuguese given names, the identification was
made by visiting the institutional web pages of each of the authors.}.

\begin{center}
\begin{tabular}{|l|l|l|l|l|l|l|l|}
\hline
\textbf{id} & \textbf{\#w} & \textbf{\#m} & \textbf{1} & \textbf{2} &
\textbf{3} & \textbf{4} & \textbf{5} \\ \hline
{\small 0001} & {\small 0} & {\small 2} & {\small 3} & {\small 29} & {\small %
0} & {\small 0} & {\small 0} \\ \hline
{\small 0002} & {\small 1} & {\small 0} & {\small 1} & {\small 0} & {\small 0%
} & {\small 0} & {\small 0} \\ \hline
{\small ...} & {\small ..} & {\small ..} & {\small ..} & {\small ..} &
{\small ..} & {\small ..} & {\small ..} \\ \hline
{\small 1,138} & {\small 2} & {\small 3} & {\small 0} & {\small 0} & {\small %
0} & {\small 0} & {\small 0} \\ \hline
\end{tabular}

{\small Table 2: Exemplifying the representation of papers in the data set (}%
$P_{(1138,5)}${\small ).}
\end{center}

There, three papers are represented: the column $id$ conveys the paper
identification, the column $\#w$ stores the number of female authors, the
column $\#m$ provides the number of male authors and the columns labeled $%
1,2,...,$ $5$ store the presence of extra subjects.

The examples in Table 2 inform that paper $0001$ has two male authors and
Business(3) and Engineering(29) as secondary (and co-occurring) subjects. It
also informs that paper $0002${\small \ }has just one female author and
Agricultural Economics(1) as its single secondary subject. The paper{\small %
\ }$1,138$ has five authors: two female and three male authors and no extra
subject.

As we aim to address interdisciplinarity issues, from the whole set of 1,138
papers we select those the have at least one extra subject. They are 535
papers whose subjects are assemble in the set $P_{535,5}^{0}.$ The
superscript $^{0}$ identifies the subset of $P_{(1138,5)}$ that comprises
all papers with at least one secondary subject. The first subscript ($535$)
indicates the size of this data set while the second subscript ($m$) stands
for the position of the extra subject in paper $i$ with $(1\leq m\leq 5)$.
There, each cell informs whether paper $i$ has subject $j$ ($p_{i,m}^{0}$ $%
=j $) with $0\leq $ $j\leq 29$. Later in the paper, the set $P_{535,5}^{0}$
is used to construct the topological representation of the 29 subjects
co-occurring with Economics in scientific publications.

\subsection{Authorship Categories}

Besides the subject concerns and depending on the authorship
characteristics, each paper belongs to at least one of the following (not
mutually exclusive) categories. The definition of the five categories of
authorship based on gender settles the basis for the identification of
patterns of research collaboration and their relation to gender. The
following list of categories is ranked in descending order of average
percentage of female authors per article: 100, 51, 42, 20 and 0,
respectively. The set papers belonging to the authorship categories are
labeled $P_{(57,5)}^{1},P_{(266,5)}^{2},P_{(209,5)}^{3},P_{(478,5)}^{4}$ and
$P_{(269,5)}^{5}$, respectively.

\begin{enumerate}
\item $P_{(57,5)}^{1}$:all authors are women (\textrm{W.Exc})

\item $P_{(266,5)}^{2}$:authors include at least one woman (\textrm{W.Inc})

\item $P_{(209,5)}^{3}$:authors include both women and men (\textrm{W\&M})

\item $P_{(478,5)}^{4}$:authors include at least one man (\textrm{M.Inc})

\item $P_{(269,5)}^{5}$:all authors are men (\textrm{M.Exc})
\end{enumerate}

Considering the articles in each category, some statistical values are
computed:

\begin{itemize}
\item the number of articles (Size)

\item the average number of authors per article ({$<$}Author%
{$>$})

\item the average percentage of female authors per article (\% female)

\item the number of articles with a single author (\#Single)

\item the average number of subjects by article ({$<$}Subject%
{$>$})

\item the number of articles with at least one extra subject (XSubject)
\end{itemize}

\subsection{Overview of the data set}

Table 3 shows the overall statistics for the 1,138 papers from 2010 to 2015
in Economics, according to the five authorship categories above presented.
While the columns correspond to the authorship categories, the rows in Table
3 provide the values obtained for the statistical indicators above described.

\begin{center}
\begin{tabular}{|c|c|c|c|c|c|c|}
\hline
{\small Authorship } & {\small All} & {\small W.Exc} & {\small W.Inc} &
{\small W\&M} & {\small M.Inc} & {\small M.Exc} \\
{\small Category} & {\small 0} & {\small 1} & {\small 2} & {\small 3} &
{\small 4} & {\small 5} \\ \hline
{\small Size} & {\small 1,138} & {\small 105} & {\small 525} & {\small 252}
& {\small 1,033} & {\small 316} \\ \hline
{\small {$<$}Author{$>$}} & {\small 2.4} & {\small 1.8} &
{\small 2.7} & {\small 3} & {\small 2.5} & {\small 2.1} \\ \hline
{\small female} & {\small 25} & {\small 100} & {\small 51} & {\small 42}
& {\small 20} & {\small 0} \\ \hline
{\small Single} & {\small 210} & {\small 46} &  &  &  & {\small 164} \\
\hline
{\small {$<$}Subject{$>$}} & {\small 2} & {\small 1.75} &
{\small 1.9} & {\small 1} & {\small 2} & {\small 2.1} \\ \hline
{\small XSubject} & {\small 535} & {\small 57} & {\small 266} & {\small 209%
} & {\small 478} & {\small 269} \\ \hline
\end{tabular}

{\small Table 3: Overall Statistics for 2010-2015 papers in Economics.}
\end{center}

The results in Table 3 seem to contradict the hypothesis that women have
more propensity to interdisciplinary research collaboration, because the
category man exclusive (M.Exc) is the one which has the higher average
number of subjects. The average number of authors is higher in the mixed
category W\&M but the woman inclusive (W.Inc) is the category with second
highest number of authors (the size of co-authorship). These results
apparently converge to the hypothesis that women prefer to work in teams.
However, this hypothesis is not confirmed by the average number of authors
of the papers in the woman exclusive category (W.Exc), being the smallest
value in the {\small {$<$}Author{$>$}} row, it indicates
that when papers are exclusively authored by women, the working teams tend
to be smaller than any of those that also include men. Looking at the number
of papers authored by a single individual (210 papers), 22\% and 78\% are
the respective percentages of female and male authorships. A similar
proportionality characterizes the percentages of woman exclusive and man
exclusive authorships (W.Exc and M.Exc) in the total amount of papers in
these two exclusive categories, they are 25\% \ (105 papers) and 75\% (316
papers), respectively.

Figures 1 and 2 show the dynamic of the five categories of authorship across
time (2010-2015). Figure 1 displays the distributions of the number of
papers in each authorship category. The plots in Figures 2 \ show the
distributions of the papers in Economics with at least one secondary subject
and the distribution of the papers in Economics with a single author.

\begin{figure}[tbh]
\begin{center}
\psfig{figure=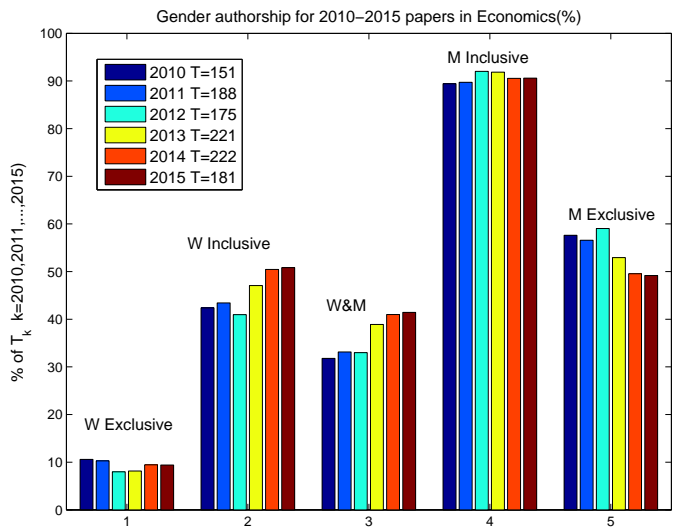,width=9truecm}
\caption{The distribution of the number of papers in each category
for papers in Economics.}
\end{center}
\end{figure}

\begin{center}
\begin{figure}[tbh]
\psfig{figure=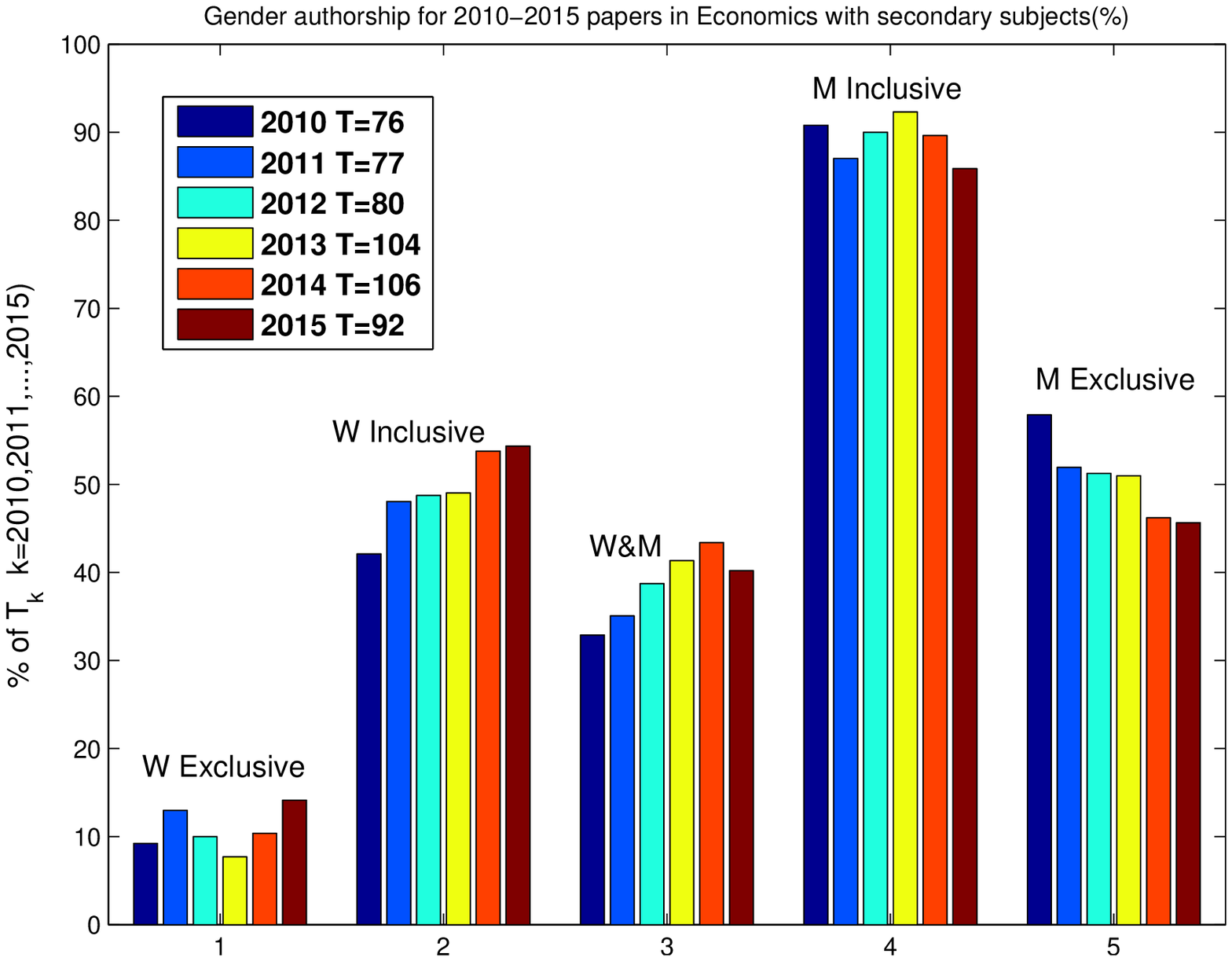,width=6truecm}
\psfig{figure=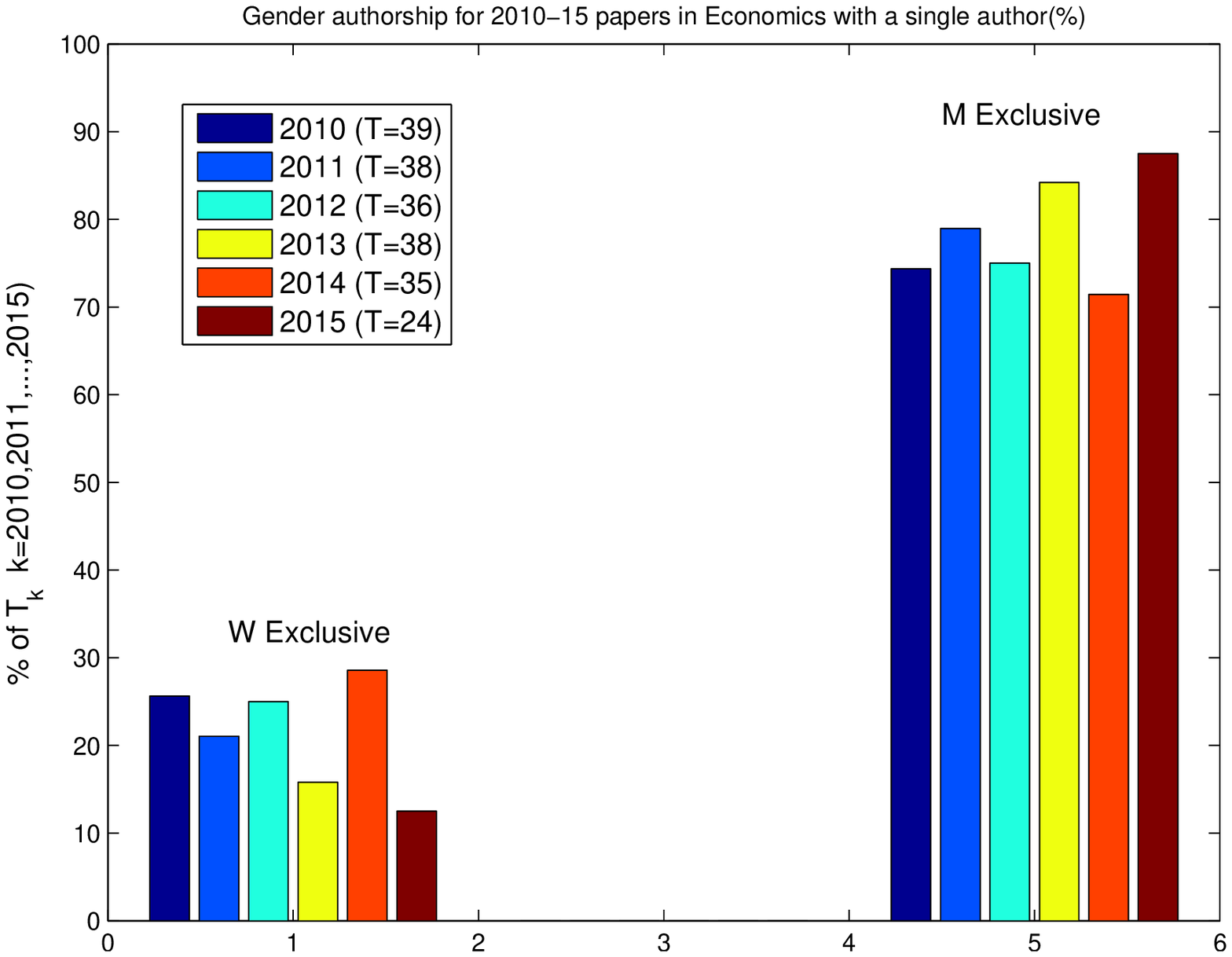,width=6truecm}
\caption{The distributions of the number of papers (a) in Economics
with at least one secondary subject and (b) in Economics with a single
author.}
\end{figure}
\end{center}

The distributions in Figures 1 and 2 are quite similar meaning that
constraining our sample to the papers with at least one extra subject does
not introduce any bias, the only (and unimportant) exception regards the man
inclusive category (M.Inc) in the first two years. The same would apply to
the distributions presented in Figure 2 if the year of 2014 was excluded. In
2014 the proportions of gender-based single authorship shows a different
balance between male and female authorships (moving from 32\% and 5\% to
25\% and 10\%, respectively). As presented in the last rows of Table 3, the
set of papers presenting a least one extra subject comprise 535 papers and
the average number of extra subjects by paper in this set is 2.

In general, the Figure 1 and Figure 2(a) reveal that there is an increasing
trend in the number of published articles across time, in all co-authorship
categories. However, there is a decrease from 2014 to 2015 with one single
exception: the papers in Economics with a secondary subject exclusively
authored by women (W.Exc).

Figures 3 shows the distributions of the relative frequencies (\%) of the
six most frequent extra subjects in each authorship category. Figure 4 shows
the distributions of the relative frequencies (\%) of the 7th to the 12th
most frequent extra subjects in each authorship category.

\begin{figure}[tbh]
\begin{center}
\psfig{figure=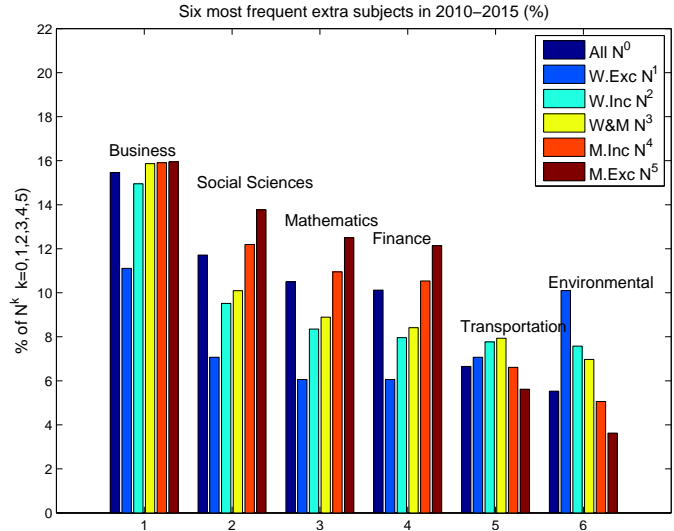,width=9truecm}
\caption{The distribution of the frequencies of the six most
frequent extra subjects in each authorship category.}
\end{center}
\end{figure}

These distributions show that the exclusive categories \textrm{W.Exc} (dark
blue) and \textrm{M.Exc} (red) display the greater fluctuations along the
different subjects. These fluctuations increase from the 5th most frequent
subject (Transportation) until the 10th (Political Sciences). The larger
imbalance between the relative frequencies of the exclusive categories
\textrm{W.Exc} and \textrm{M.Exc} relies on the subjects Environmental
Sciences, Management and Political Sciences. When compared with the high
homogeneous distribution that characterizes Business, the relative
frequencies of Environmental Sciences, Management and Political Sciences
increase in the woman exclusive category (\textrm{W.Exc}) in the same
proportion they decrease in the man exclusive (\textrm{M.Exc)} one. These
very first results indicate that the subjects Environmental Sciences,
Management and Political Sciences are more likely to co-occur in
female-dominated papers in Economics.

\begin{figure}[tbh]
\begin{center}
\psfig{figure=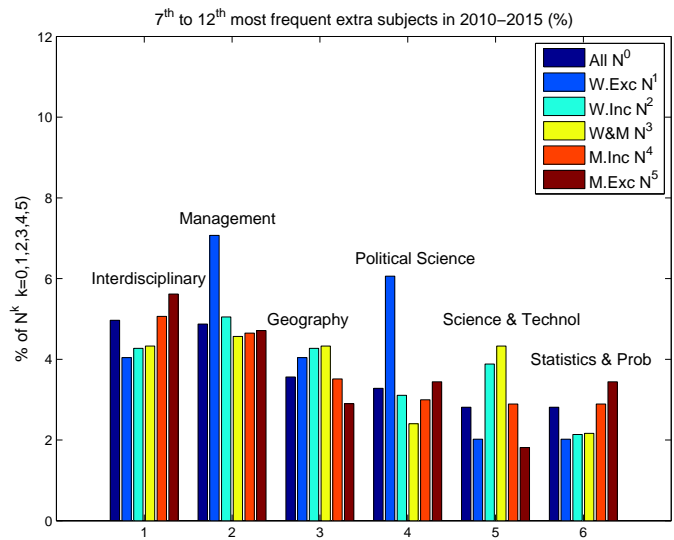,width=9truecm}
\caption{The distribution of the frequencies (\%) of the seventh to
the twelfth most frequent extra subjects in each authorship category.}
\end{center}
\end{figure}

In summary, when considering papers published in WoS indexed journals over
the period 2010-2015 in the scientific domain of Economics and whose authors
are affiliated to a Portuguese institution, our results suggest that:

\begin{enumerate}
\item men have more propensity to interdisciplinary research collaboration,
since the man exclusive category ($\mathrm{M.Exc)}$ has the highest average
number of subjects

\item the woman inclusive (\textrm{W.Inc}) is the category with the second
highest average number of authors. These results apparently converge to the
hypothesis that women prefer to work in teams but

\item when papers are exclusively authored by women (\textrm{W.Exc}), the
working teams tend to be smaller than any of those that also include men

\item academic women compared with their male counterparts reveal preference
for the subjects Environmental Sciences, Management and Political Sciences

\item conversely, the subjects Social Sciences, Mathematics and Finance
display higher frequencies in papers either inclusively (\textrm{M.Inc}) or
exclusively authored by men (\textrm{M.Exc})
\end{enumerate}

In the next section, a network approach is applied to combine the gender
authorship perspective with the analysis of interdisciplinarity. To this
end, the categories of articles are used to construct the topological
representation of the 29 subjects (Table 1) co-occurring with Economics in
scientific publications.

\section{Network Induction}

Network induction makes reference to the method by which networks are
created on the basis of a certain data set or system. As earlier mentioned,
network approaches are quite common in the analysis of systems where a
network representation is the most intuitive. As connecting the elementary
units of a system may be conceived in many different ways, that choice may
depend strongly on the available empirical data and on the questions that a
network analysis aims to address. Here, six bipartite networks are induced
from the subsets of papers defined by the authorship categories presented in
Section 2.1.

The frequency of co-occurrence of each pair of subjects defines the
existence of every link in the networks by authorship category. They are
weighted graphs since the weight of each link corresponds to the frequency
of co-occurrence of the linked pair of subjects. In the next section, those
weighted networks are further analyzed through the construction of their
corresponding minimal spanning trees (MST). In so doing, we are able to
emphasize the main topological patterns that emerge from each network
representation and to discuss their interpretation and relation to gender.

\subsection{Bipartite Graphs}

A bipartite network $N$ consists of two partitions of nodes $V$ and $W$,
such that edges connect nodes from different partitions, but never those in
the same partition. A one-mode projection of such a bipartite network onto $%
V $ is a network consisting of the nodes in $V$; two nodes $v$ and $v\prime $
are connected in the one-mode projection, if and only if there exist a node $%
w\in W$ such that $(v,w)$ and $(v\prime ,w)$ are edges in the corresponding
bipartite network ($N$). In the following, we explore six bipartite networks
and their corresponding one-mode projections.

\subsection{Connecting subjects}

Each bipartite network by authorship category consists of the following
partitions:

\begin{itemize}
\item the set $S$ of 29 subjects presented in Table 1 and

\item one set of papers ($P^{k}$) by authorship category ($k=\{0,1,2,3,4,5\}$%
) presented in Section 2.1.
\end{itemize}

In the each network ($N^{k}$), two subjects are linked if and only if they
co-occur in at least one paper of $P^{k}$, having each paper at most five
subjects. Therefore, the links in each network $(N^{k})$ are weighted by the
number of coincident papers a pair of subjects share in $P^{k}$.
Consequently, every link $L_{(i,j)}^{k}$ in $N^{k}$ takes value in the set \{ 0,1,2,..., size($P^{k}$)\}.

As an example and considering that in $P_{(535,5)}^{0}$(the authorship
category comprising all papers with at least one secondary subject) there
are just three papers where the subjects Agricultural Economics and Finance
co-occur yields $L_{(1,8)}^{0}=3$. Another example is $L_{(1,3)}^{0}=1$ due
to the mutual single co-occurrence of Agricultural Economics and Business in
$P_{(535,5)}^{0}$. Among the many examples of missing links there are the
cases of and Education and Finance($L_{(6,8)}^{0}=0$) since these two
subject do not co-occur in any paper of $P_{(535,5)}^{0}$.

Having induced the networks ($N^{k}$) for each authorship category, we are
able to have a complete representation of the relationship among the
subjects co-occurring in each authorship category defined based on gender.
However, it so happens that neither the densely-connected nature nor the
existence of disconnected components of these networks helps to find out
whether there is a dominant pattern in the structure of subjects. The large
number of links make the extraction of the truly relevant connections
forming the network a challenging problem. One first step in the direction
of extracting relevant information from the networks may be targeted at
obtaining the corresponding MST(\cite{Arau03};\cite{Arau07};\cite{Arau12};%
\cite{Arau16b}).

\subsubsection{From complete networks to minimum spanning trees}

In the construction of a MST by the \textit{nearest neighbor} method, one
defines the subjects (in Table 1) as the nodes ($n_{i}^{k}$) of a weighted
and connected\footnote{%
The hierarchical clustering process considers just the largest connected
component of each network ($N^{k}$). Therefore, depending on the authorship
category ($k$) the resulting MSTs have different sizes, as indicated in the
first row of Table 4.} network ($N^{k}$) where the distance $d_{ij}^{k}$
between each pair of subjects $i$ and $j$ corresponds to the inverse of
weight of the link ($d_{ij}^{k}=\frac{1}{L_{ij}^{k}}$) between $i$ and $j$.

From the $nxn$ distance matrix $D_{i,j}^{k}$, a hierarchical clustering is
performed using the \textit{nearest neighbor} method. Initially $n$ clusters
corresponding to the $n$ subjects are considered. Then, at each step, two
clusters $c_{i}$ and $c_{j}$ are clumped into a single cluster if

\begin{center}
$d^{k}\{c_{i},c_{j}\}=\min \{d^{k}\{c_{i},c_{j}\}\}$
\end{center}

with the distance between clusters being defined by

\begin{center}
$d^{k}\{c_{i},c_{j}\}=\min \{d_{pq}^{k}\}$ with $p\in c_{i}$ and $q\in c_{j}$
\end{center}

This process is continued until there is a single cluster. This clustering
process is also known as the \textit{single link method}, being the method
by which one obtains the MST of a graph \cite{Arau00}.

In a connected graph, the MST is a tree of $n-1$ edges that minimizes the
sum of the edge distances. In a network with $n$ nodes, the hierarchical
clustering process takes $n-1$ steps to be completed, and uses, at each
step, a particular distance $d_{i,j}^{k}$ $\in $ $D^{k}$ to clump two
clusters into a single one.

Let $C=\{d_{q}\},q=1,...,N-1$, be the set of distances $d_{i,j}^{k}$ $\in $ $%
D^{k}$ used at each step of the clustering, and $thr=\max \{d_{q}\}$. It
follows that $thr=d_{N-1}^{k}$.

The result of the hierarchical clustering process leading to the MST is
usually described by means of a dendrogram. During this process, a unique
color is assigned to each group of nodes within the dendrogram whose linkage
is less than $T$ times the value of the threshold distance $(thr$). In the
dendrogram presented here, $T$ is set to $1.2$.

Six clusters can be observed in the dendrogram of $N^{0}$(the network of the
authorship category comprising all papers with at least one secondary
subject) as Figure 5 shows. The colors assigned to these clusters will be
hereafter used in the identification of the same partitions of subjects
whenever represented in a MST.

The dendrogram in Figure 5 shows that the subjects Hospitality and Leisure,
Sports \& Tourism are the first to be clumped since their occurrences are
perfectly correlated in $P_{(535,5)}^{0}$. On the other hand, the papers on
these two subjects remain almost isolated from any other subject matter in
the overall set of papers being considered. The next cluster being defined
comprises the subjects Business and Finance (colored blue). Being followed
by the large cluster including Mathematics, Statistics, Social Sciences and
Interdisciplinary Sciences (yellow). Another early defined cluster clumps
together Transportation, Operational Research, Engineering and Science \&
Technology (turquoise). Further analyzing a dendrogram by its corresponding
MST allows for observing the extent to which clusters give place to branches
on the tree and whether different motifs emerge from the clusters
positioning on the trees.

\begin{figure}[tbh]
\begin{center}
\psfig{figure=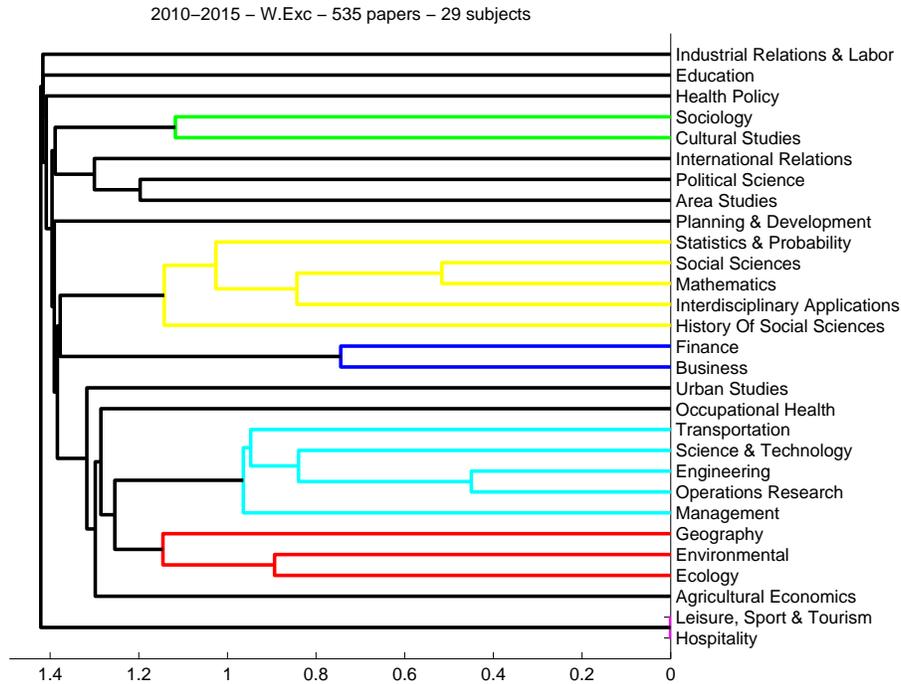,width=12truecm}
\caption{The dendrogram shows the hierarchical clustering process
applied to $N^{0}$.}
\end{center}
\end{figure}

Figure 6 shows the representation of the corresponding MST. It is worth
noting that closeness on the MST depends on the connection strength (the
weight of the links) in{\small \ }$N^{0}$, meaning that when two subjects
co-occur in many papers of $P_{(535,5)}^{0}$ (being therefore strongly
connected) they occupy close positions on this tree.

While the dendrograms provide information on the distances at which the
subjects are clumped into clusters, their corresponding minimum spanning
trees allow for the identification of at least four important aspects that
are not directly stated in the dendrograms.

\begin{enumerate}
\item Branches: the way nodes organize themselves in different ramifications
of the tree

\item Motifs: the prevalence of \emph{star} motifs and/or\emph{\ path}
motifs in the tree

\item Connectivity: highly connected and weakly connected nodes

\item Centrality: the nodes occupying highly central positions and,
conversely, those occupying the leafs of the tree
\end{enumerate}

\begin{figure}[h]
\begin{center}
\psfig{figure=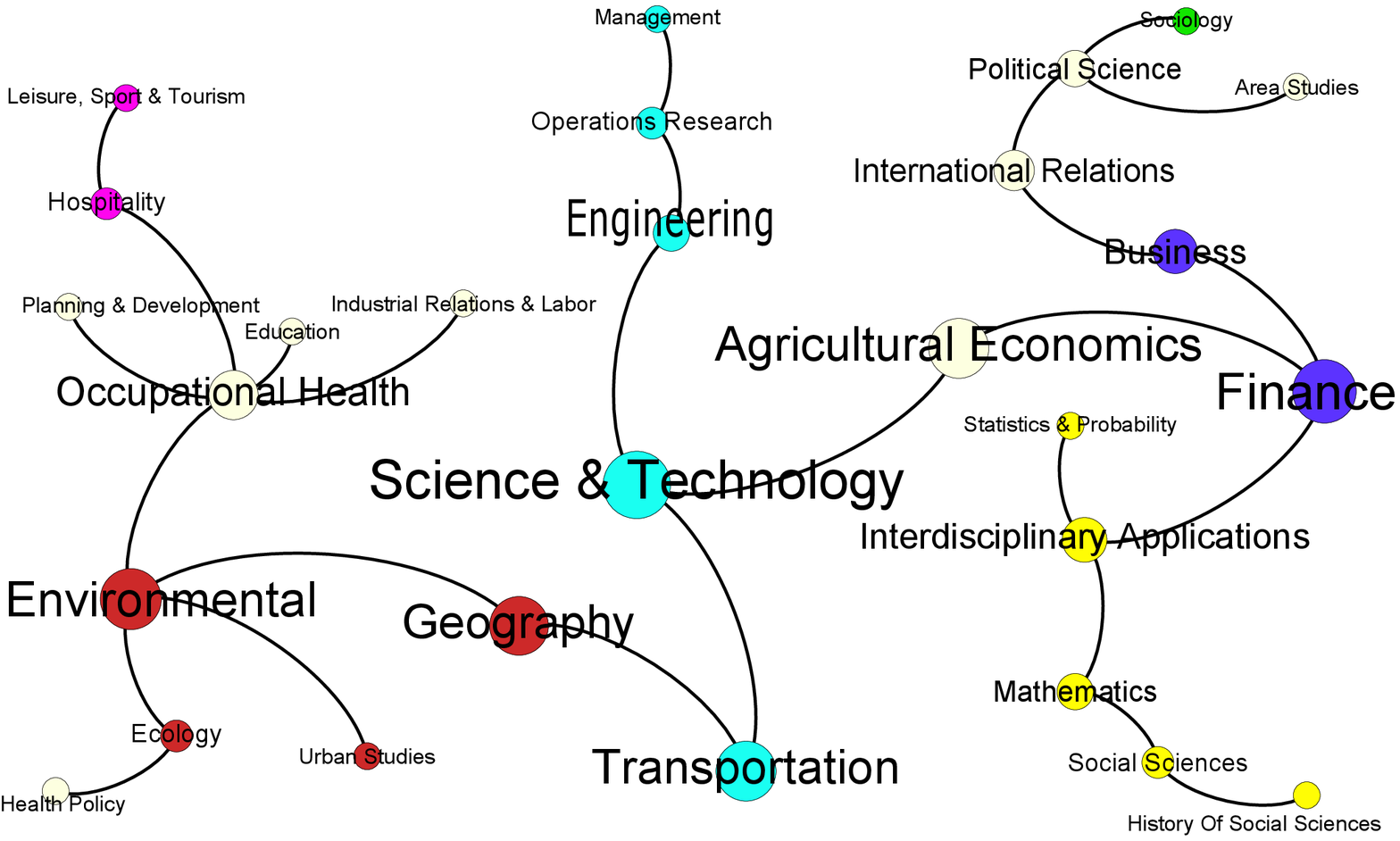,width=14truecm}
\caption{The MST of $N^0$, which comprises all papers with at least one secondary subject.}
\end{center}
\end{figure}

The observation of the MST presented in Figure 6 suggests that besides a
"core"\ cluster, there are at least three
important branches separating "classical"\ %
subjects; "technological"\ subjects and
"environment-related"\ subjects. Figure 6 also
shows that there are two highly connected nodes: Environmental Sciences and
Cultural Studies.

In what concerns centrality, the subjects Political Sciences and Science \&
Technology occupy positions of great centrality on the tree. A distinct
situation characterizes Education and Industrial Relations \& Labor which
occupy leaf positions on the MST. These two subjects, together with the
cluster that joins Hospitality and Leisure, Sports \& Tourism are the last
ones to be connected in the hierarchical clustering process, as the
dendrogram of Figure 5 shows.

\subsubsection{The minimum spanning trees by authorship category}

Since we hypothesized that specific characteristics could come out and shape
the structures of the networks of subjects and that these characteristics
may be associated to some ordering emerging from gender, here we consider
the subsets of papers defined by the authorship categories $%
P_{(57,5)}^{1},P_{(266,5)}^{2},P_{(209,5)}^{3},P_{(478,5)}^{4}$ and $%
P_{(269,5)}^{5}$. In applying the hierarchical clustering process to each
subset provides the following MSTs. They are ranked in descending of average
percentage of female authors per article (as in Section 2.1).

\begin{enumerate}
\item All authors are women (\textrm{W.Exc-MST})

\item Authors include at least one woman (\textrm{W.Inc-MST})

\item Authors include both women and men (\textrm{W\&M-MST})

\item Authors include at least one man (\textrm{M.Inc-MST})

\item All authors are men (\textrm{M.Exc-MST})
\end{enumerate}

Obtaining the MST of a given network implies that the network is connected.
Therefore, the application of the hierarchical clustering process to each
network ($N^{k}$) by authorship category considers just the largest
connected component of each network. Thereafter, depending on the authorship
category ($k$) the resulting MSTs have different sizes, which are indicated
in the first row of Table 4.($N^{0}$)

Figures 7 and 8 present the minimum spanning trees of the gender exclusive
authorship categories ({\small W.Exc-MST }and {\small M.Exc-MST}), being the
nodes colored according to the partitions of subjects as defined in the
dendrogram presented in Figure 5.

\begin{figure}[h]
\begin{center}
\psfig{figure=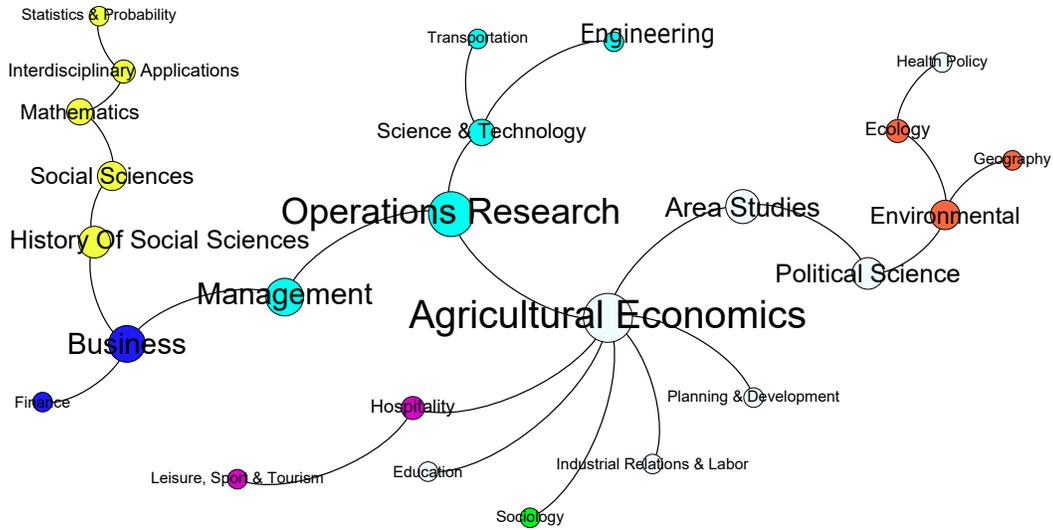,width=14truecm}
\caption{The MST of the woman exclusive category (W.Exc-MST).}
\end{center}
\end{figure}

\begin{figure}[h]
\begin{center}
\psfig{figure=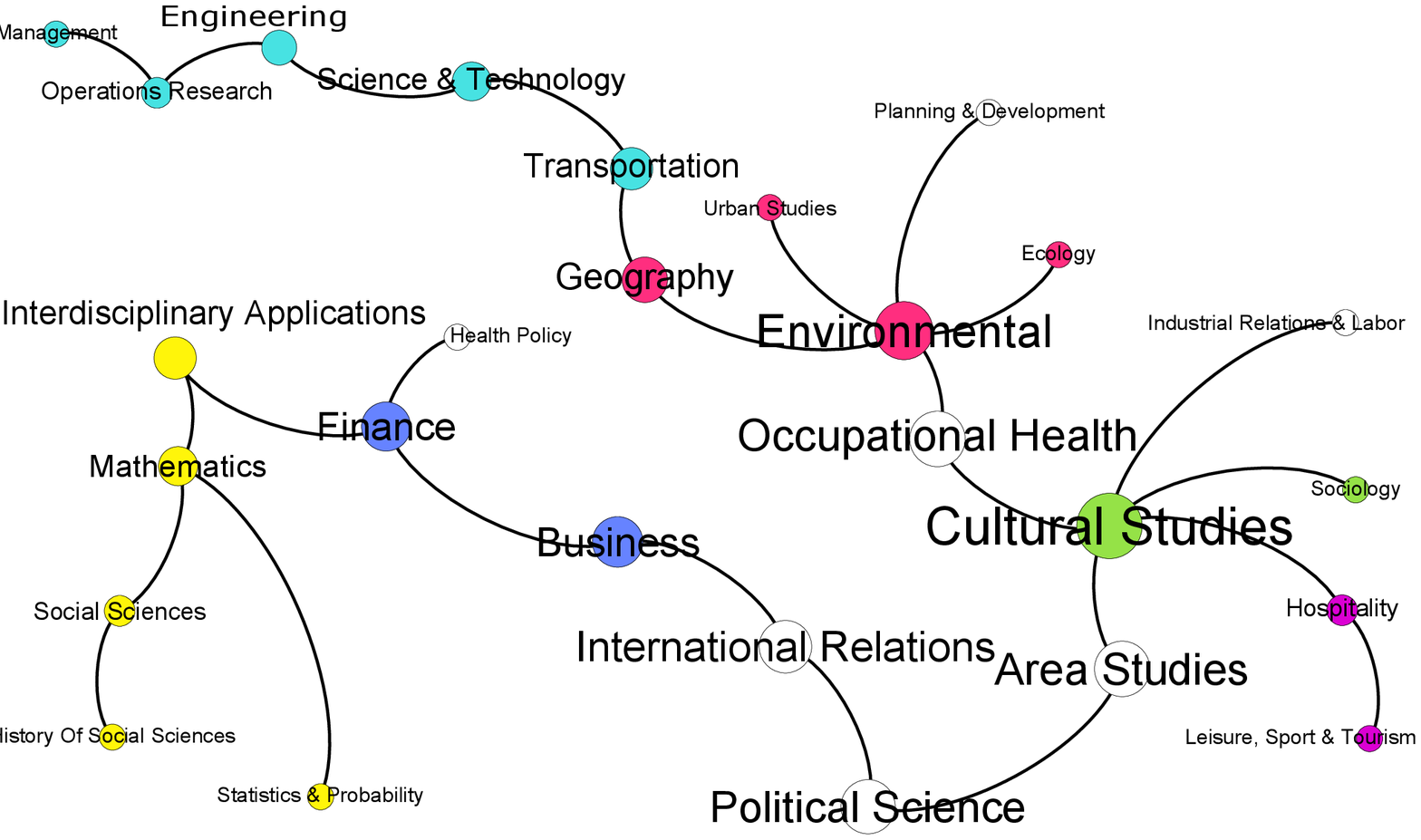,width=14truecm}
\caption{The MST of the man exclusive category (M.Exc-MST).}
\end{center}
\end{figure}

These networks are quite similar in the way nodes organize themselves in
different branches (clusters) on the tree. However, there is an important
difference concerning the centrality of certain nodes and the positioning of
the main branches on the trees.

When centrality matters, Management occupies a central position in the woman
exclusive ({\small W.Exc-MST in Figure 7}) but looses centrality in the man
exclusive one ({\small M.Exc-MST in Figure 8}).
The positioning of the
"core", "classical" and "technological" branches
suffer important changes when compared to their situation in the global
{\small MST (}${\small N}^{0}$ in {\small Figure 6)}. While the
"core"\ and the "classical" branches remain linked in both the female and the man
exclusive, the "technological" and the
"core" branches, that in the global {\small MST}
were linked through the Agricultural node are far away in the man exclusive
MST {\small (M.Exc-MST)}. The fact that they occupy close positions on the
woman exclusive MST ({\small W.Exc-MST}) is probably associated to the
greater centrality of the subject Management in this tree.

The increase of centrality of the subject Management in the woman exclusive
{\small MST} together with the presence of the subject Agricultural
Economics has an important bearing on that tree ({\small W.Exc-MST)},
showing that when papers authorship includes just women, the larger
distances between subjects in the network tend to be reduced due to an
important increase in the relative number of papers having Management as a
secondary subject.

\subsubsection{Tree motifs}

The adoption of a network approach provides well-known notions of graph
theory to fully characterize the structure of the networks. Here, and since
our analysis relies on the minimum spanning trees, we concentrate on the
calculation of just two topological coefficients, both measured at the
network level.

\begin{figure}[tbh]
\begin{center}
\psfig{figure=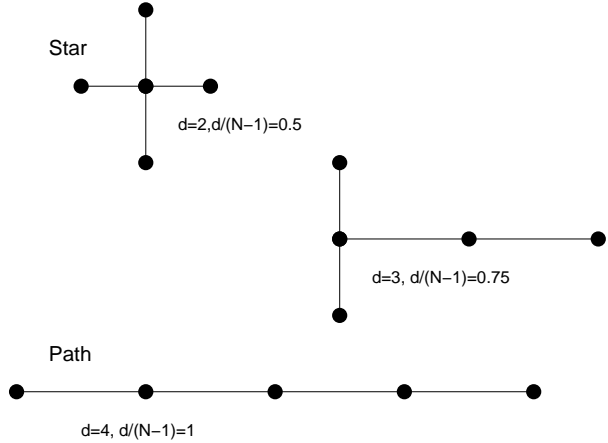,width=8truecm}
\caption{Examples of different motifs of a tree with five nodes:
from a Star to a Path motif.}
\end{center}
\end{figure}

The first one is the number of leafs $(l)$ in the MST, i.e., the number of
nodes with degree one. The second coefficient is the MST diameter ($d$),
measuring the shortest distance between the two most distant nodes on the
tree. The choice of these coefficients allows to characterize tree motifs
with different shapes: from a pure \emph{star} to a pure \emph{path} motif.
Figure 9 shows examples of different motifs occurring on a tree of just five
nodes ($N=5$) and the values of each corresponding diameter ($d$) and
coefficient $\frac{d}{N-1}.$

It so happens that when the number of nodes of the tree is greater than 2,
and depending on the motif that the MST approaches, its diameter ranges in
between $2$ and $N-1$ ($2\leq d\leq N-1$). The closer is $\frac{d}{N-1}$to
1, the smaller is the similarity of the MST to a \emph{star} motif.
Moreover, the number of leafs ranges in between exactly the same values but
in the opposite direction, the closer $l$ is to 1, the smaller is the
similarity of the MST to a \emph{path} motif.

Table 4 shows the values of $N,d,l$ and $\frac{d}{N-1}$ computed for the
five trees by the authorship category. The first row in Table 4 displays the
size of each MST, i.e., the number of nodes in each MST. The last row shows
the values obtained for the coefficient $\frac{d}{N-1}$, which are limited
between $\frac{2}{N-1}$(\emph{star}) and $1$ (\emph{path}).

\begin{center}
\begin{tabular}{|l|c|c|c|c|c|}
\hline
{\small Authorship Category} & {\small 1} & {\small 2} & \multicolumn{1}{|c|}%
{\small 3} & {\small 4} & {\small 5} \\
\multicolumn{1}{|c|}{\small MST} & {\small W.Exc} & {\small W.Inc} & {\small %
W\&M} & {\small M.Inc} & {\small M.Exc} \\ \hline
\multicolumn{1}{|c|}{${\small N}$} & {\small 25} & {\small 28} &
\multicolumn{1}{|c|}{\small 27} & {\small 29} & {\small 27} \\ \hline
\multicolumn{1}{|c|}{${\small d}$} & {\small 13} & {\small 12} &
\multicolumn{1}{|c|}{\small 11} & {\small 12} & {\small 17} \\ \hline
\multicolumn{1}{|c|}{$l$} & {\small 11} & {\small 11} & \multicolumn{1}{|c|}%
{\small 12} & {\small 13} & {\small 10} \\ \hline
\multicolumn{1}{|c|}{$\frac{d}{N-1}$} & {\small 0.54} & {\small 0.44} &
\multicolumn{1}{|c|}{\small 0.44} & {\small 0.43} & {\small 0.65} \\ \hline
\multicolumn{1}{|c|}{\small \% female authors} & {\small 100} & {\small 51}
& {\small 42} & {\small 23} & {\small 0} \\ \hline
\end{tabular}

{\small Table 4: Topological coefficients computed from the MST of each
authorship category.}
\end{center}

Although the five networks have similar sizes, there is a remarkable
difference in the values obtained for the man exclusive tree ({\small %
M.Exc-MST}). When women are excluded, the network of subjects displays an
much higher diameter ($d$), showing large distances among subjects are
enlarged. The also important decrease in the number of leafs ($l$) indicates
that this network develops a entirely different structure when compared with
the other MSTs by authorship category.

The plots in Figure 10 show the number of leafs ($l)$ , the diameter ($d)$
and the ratio $\frac{|d-l|}{N}$ across the different categories of
authorship. As, depending on the specific tree motif, the values of $d$ and $%
l$ move in opposite directions, in computing the absolute value of the
difference $d-l$ relative to $N$ helps to emphasize the distinguish
structure of the MST that characterizes the man exclusive network ({\small %
M.Exc-MST}).

\begin{figure}[tbh]
\begin{center}
\psfig{figure=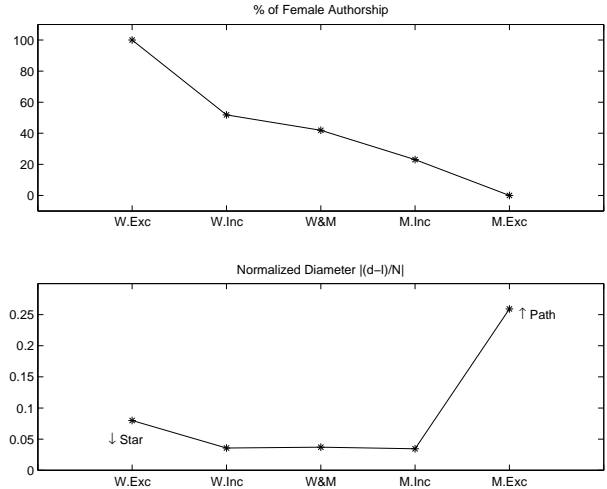,width=8truecm}
\caption{The \% of female authorship along with the
different MST categories and the corresponding evolution of $\frac{|d-l|}{N}
$.}
\end{center}
\end{figure}

In the broader set of papers published in WoS indexed journals over the
period 2010-2015 in the scientific domain of Economics and having at least
one author affiliated to a Portuguese institution, as the percentage of
female authorship decreases, the MST obtained from the corresponding network
of subjects moves from a star configuration to a path motif. In so doing,
the larger distances between subjects are enlarged and the number of poorly
connected subjects increases. If, conversely, the network of subjects has a
high percentage of female authorship, the corresponding MST approaches a
star motif, the number of leafs is enlarged and the corresponding diameter
decreases\footnote{%
The woman exclusive authorship ({\small W.Exc-MST}) shows a small deviation
in the opposite direction. However, the fact that this network was induced
from a small number of papers (57) might introduce some bias in its shape.}.

\section{Conclusion}

There are many ways to link the elementary units of system in order to
induce a network. Choosing the most suitable way depends strongly on the
available empirical data and on the research questions that a network
analysis aims to address. Regarding available empirical data, most of
bibliometric databases have a strong weakness concerning the study of the
differences by gender. In what concerns research questions, gender
differences in collaborative research and interdisciplinarity in scientific
outputs have received little attention when compared with the growing
importance that women hold in academia and research.

From the set of papers published in WoS indexed journals over the period
2010-2015 in the scientific domain of Economics and having at least one
author affiliated to a Portuguese institution, our results apparently
converge to the hypothesis that women prefer to work in teams. However, they
also indicate that when papers are exclusively authored by women, the
working teams tend to be smaller than any of those that also include men.
These results converge to the mixed results reported in the literature,
where different units of analysis, measures, methods and samples were
adopted (\cite{Abramo15};\cite{McDo83}; \cite{Meng16}-\cite{Uhly15}).

Regarding interdisciplinarity, our findings seem to contradict the
hypothesis that women have more propensity to interdisciplinary research
collaboration \cite{Abramo15}. Moreover, we found that academic women in
Economics compared with their male counterparts reveal preference for the
subjects Environmental Sciences, Management and Political Sciences and that,
conversely, the subjects Social Sciences, Mathematics and Finance display
higher frequencies in papers either inclusively or exclusively authored by
men.

Our main contribution relies in the adoption of a network approach allowing
to uncover the emergence of a specific pattern when the network of
scientific subjects is induced from a set of papers exclusively authored by
men. Such a male exclusive authorship condition is found to be the solely
responsible for the emergence of that specific shape in the structure of the
network.

Moving away from a \emph{star} motif together with the loss of centrality of
the subject Management have an important bearing on the structure of the%
{\small \ }male exclusive authorship network: when papers authorship
includes just men, the larger distances between subjects in the network
become even larger and this is mainly due to a decrease in the relative
number of papers having Management as a secondary subject. We find enough
evidence that gender imbalance in scientific authorships brings a peculiar
trait to the networks of subjects. Such a peculiar trait might facilitate
future network analyses of research collaboration and interdisciplinarity.

Acknowledgement:

The research reported in this paper was developed under the PLOTINA project ("Promoting gender balance and inclusion in research, innovation and training"), which has received funding from the  European Union's Horizon 2020 research and innovation programme, under Grant Agreement No 666008 (www.plotina.eu). The views and opinions expressed in this publication are the sole responsibility of the author(s) and do not necessarily reflect the views of the European Commission.
Financial support by FCT (Funda\c{c}\~{a}o
para a Ci\^{e}ncia e a Tecnologia), Portugal is gratefully acknowledged.
This article is part of the Strategic Project: UID/ECO/00436/2013.
The authors thank Marilei Kroetz for her
work in the gender identification process.

\end{document}